\documentclass[12pt]{iopart}

\usepackage{graphicx}

\usepackage{iopams}

\usepackage{subfigure}

\newcommand{\mean}[1]{\langle #1 \rangle}
\newcommand{\PI}{ \mathrm P_{\mathrm i}}
\newcommand{\weq}{k^{\mathrm{eq}}}
\newcommand{\wplus}{k^{+}}
\newcommand{\wminus}{k^{-}}

\newcommand{\thetap}{\theta_{+}}
\newcommand{\thetam}{\theta_{-}}
\newcommand{\corr}[2]{\langle #1 #2 \rangle}
\newcommand{\n}{n}
\newcommand{\x}{x}
\newcommand{\y}{y}
\newcommand{\dn}{d}
\newcommand{\fk}{\kappa}
\newcommand{\cv}{C_{\dot{x}}}
\newcommand{\res}{R_{\dot{x}}}

\newcommand{\ftcv}{\tilde{C}_{\dot{x}}}
\newcommand{\ftres}{\tilde{R}_{\dot{x}}}

\newcommand{\qk}{Q_{\mathrm{P}}}

\newcommand{\qdotk}{\dot{Q}_{\mathrm{P}}}
\newcommand{\qdotm}{\dot{Q}_{\mathrm{M}}}

\newcommand{\qdoths}{\dot{Q}^{\mathrm{HS}}}

\newcommand{\fnull}{\mathrm{F}_{\mathrm{o}}}
\newcommand{\feins}{\mathrm{F}_{1}}

\newcommand{\fext}{f_{\mathrm{ex}}}
\newcommand{\dmu}{\Delta\mu}

\newcommand{\etaq}{\eta_{\mathrm{Q}}}
\newcommand{\etastk}{\eta_{\mathrm{S}}}
\newcommand{\wgplus}{\bar{k}^{+}}
\newcommand{\wgminus}{\bar{k}^{-}}
\newcommand{\kt}{k_{\mathrm{B}}T}

\begin{document}

\title{Efficiencies of a molecular motor:\\ A generic hybrid model applied to the $\feins$--ATPase}

\author{Eva Zimmermann and Udo Seifert}

\address{II. Institut f\"ur Theoretische Physik, Universit\"at Stuttgart, 70550 Stuttgart, Germany}

\begin{abstract}
In a single molecule assay, the motion of a molecular motor
is often inferred from measuring the stochastic trajectory
of a large probe particle attached to it. We discuss a simple
model for  this generic set-up taking into
account explicitly the elastic coupling between probe and motor.
The combined dynamics consists of discrete steps of the motor and
continuous Brownian motion of the probe. Motivated by recent experiments
on the $\feins$--ATPase, we investigate three types of efficiencies both
in simulations and a Gaussian approximation. Overall, we obtain good
quantitative agreement with the experimental data. In
particular, we clarify  the conditions under which one of these efficiencies
becomes larger than one. 
\end{abstract}


\section{Introduction}
Molecular motors are protein complexes of the size of nanometers that convert chemical energy into mechanical motion \cite{howard,schl03}. Operating in an aqueous solution they exhibit stochastic dynamics and energetics due to the influence of thermal fluctuations. Unbalanced concentrations of the molecules providing  chemical energy as input cause the motor proteins to operate under nonequilibrium conditions which induces a rectified motion with non--zero average velocity. Consequently, molecular motors are often modelled using Langevin, Fokker--Planck or master equations. The so called ratchet models combine continuous diffusive spatial motion with stochastic switching between different potentials corresponding to different chemical states \cite{juel97,reim02a}. Alternatively, transitions among a discrete state space governed by master equations provide another possibility to model molecular motors \cite{kolo07,lau07a,liep07a,lipo09a,astu10}.

A quantity of general interest is the efficiency of such stochastic machines \cite{parm99,parr02,seif11a,efre11,kawa11}. For motor proteins, different kinds of efficiencies can be defined  depending on whether one focuses on the work against an external force, i.e., thermodynamic efficiency or whether work against viscous friction is also taken into account like in the Stokes or generalized efficiency \cite{dere99,wang02a,suzu03,wang05a,qian08,boks09}.

Experimentally, the properties of motor proteins can be investigated in single molecule experiments by attaching probe particles of the size of micrometers to the motor protein and by observing the trajectories of the probes. Additionally, such probes allow to exert forces on these motor proteins \cite{noji97,rock01,itoh04,cart05}. Literally speaking, in these assays one cannot observe the motion of the motor directly but rather has to infer its properties from analyzing the trajectory of the probe particle. Generically, some elastic linker couples these two elements. Inferring properties of the motor protein requires to consider the interaction effects that depend on the linkage between motor protein and probe \cite{pesk00,pesk00a,chen00,chen02,zeld05,wang08,kunw08,bouz10}. 

In the present paper, we discuss a minimal hybrid model for such a motor protein assay that includes this elastic link explicitly. The motor protein and the probe will be modelled as two degrees of freedom moving along a spatial coordinate. In particular, we investigate different kinds of efficiencies used previously to describe the energetics of molecular motors and compare our results quantitatively to recent experiments of the rotary motor protein $\feins$--ATPase \cite{toya10,toya11,toya11a,haya10}. Previous theoretical modelling of $\feins$--ATPase using a discrete state model as well as a ratchet model assuming the probe to stick directly at the motor has especially focussed on the dependence of the rotational behaviour on friction, external forces, nucleotide concentrations and temperature as well as on chemical and thermodynamic efficiency and the fluctuation theorem \cite{gasp07,gerr10}. Detailed modelling of the rotary mechanism and the involved subunits can be found in \cite{wang98,sun04}.

This particularly well studied molecular motor consists of three $\alpha$ and three $\beta$ subunits arranged around a central $\gamma$ shaft \cite{okun11}. Binding and hydrolysis of an ATP molecule at a $\beta$ subunit drives a rotation of the $\gamma$ shaft of $120^\circ$ \cite{noji97} which has been observed to consist of two substeps of $90^\circ$ and $30^\circ$ \cite{yasu01}. An external torque exerted on the $\gamma$ shaft (as experimentally done in \cite{itoh04} or by the $\fnull$ part within the cell) induces ATP synthesis. Coupled to the membrane embedded $\fnull$ part, $\feins$--ATPase provides ATP for further hydrolysis reactions therefore being an important part in the energy transfer of cells. Experimental observations of the $\feins$--ATPase in the hydrolysis direction include the measurements of different kinds of efficiencies. The Stokes efficiency, a Stokes efficiency confined to single jumping events and the thermodynamic efficiency, especially at stall conditions, have been investigated \cite{yasu98,mune07,toya11}. These experiments led to values for the Stokes efficiency and the thermodynamic efficiency of almost 1 suggesting that the $\feins$--ATPase can use almost the complete chemical energy either to drive the probe through a viscous medium or to perform work against an external force. Recently, a measure of the efficiency that takes explicitely care of fluctuations was introduced \cite{toya10}. The definition of efficiency used there also provided values close to 1 for the examined parameters. Our analysis will show that the latter efficiency can easily reach values larger than 1.

\section{Hybrid--Model}
\subsection{Single molecule dynamics}
\begin{figure}[top]
 \centering
\includegraphics[width=12cm]{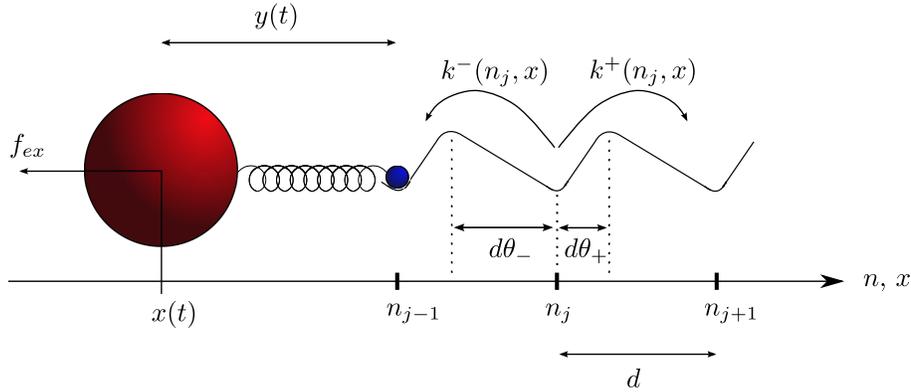}
\caption{Schematic representation of the motor protein (blue) with attached probe (red). The instantaneous distance between motor protein and probe is denoted by $y(t)$. The probe moves along a continuous spatial coordinate $x(t)$ an is subject to an external force $\fext$. With transition rates $k^{\pm}(n_j,x)$ the motor protein jumps at times $\tau_j$ between discrete states $n_j$ separated a distance $d$. The load sharing factors $\thetap$ and $\thetam$ indicate the position of an underlying unresolved potential barrier relative to the potential minimum.}
\label{Model}
\end{figure}

The one--dimensional model we will use to describe a molecular motor with an attached probe particle consists of two degrees of freedom representing the motor protein at position $\n(t)$ and the probe at position $\x(t)$, respectively, see figure \ref{Model}. For a rotary motor like the $\feins$--ATPase, the rotary motion is mapped to a linear one for simplicity.
Both constituents are linked via a harmonic potential
\begin{equation}
 V(n,x)=\frac{\fk}{2}(n-x)^2
\end{equation}
with spring constant $\fk$, where we have included a possible rest length of the linker into the definition of $x$.

The motion of the probe particle is described by an overdamped Langevin equation with friction coefficient $\gamma$ and constant external force $\fext$,
\begin{equation}
 \dot{\x}(t) =(-\frac{\partial V}{\partial x}-\fext)/\gamma+\zeta(t), 
 \label{LG}
\end{equation}
including the random force $\gamma\zeta(t)$ that the solvent exerts on the probe. The thermal fluctuations are assumed to be Gaussian white noise with zero mean and correlations $\corr{\zeta(t_1)}{\zeta(t_2)} = 2 \kt\delta(t_1-t_2)/\gamma$ where $k_{\mathrm{B}}$ is Boltzmann's constant and $T$ the temperature of the solvent.

The motor protein jumps at times $\tau_j$ from $\n_j$ to $\n_j\pm d$ with transition rates $k^{\pm}(n_j,x)$, hydrolyzing (or synthesizing) one ATP molecule per jump which corresponds to tight mechanochemical coupling. In this minimal model, we take into account only one chemical state. The transition rates have to fulfill a local detailed balance condition of the form
\begin{equation}
 \frac{\wplus(n,x)}{\wminus(n+d,x)}=\exp[(\Delta\mu-V(\n+d,\x)+V(\n,\x))/\kt]
\label{ldb}
\end{equation}
where we assume that the jumps of the motor protein take place instantaneously. The free energy change of the solvent
\begin{equation}
\Delta\mu\equiv\mu_{\mathrm{ATP}}-\mu_{\mathrm{ADP}}-\mu_{\PI}
\end{equation}
is associated with ATP turnover. Implementing mass action law kinetics and the concept of a barrier in the potential of mean force for the unresolved chemical steps, the individual rates become
\begin{eqnarray}
 \fl\wplus(\n,\x) &= \weq \exp [\Delta\mu_{\mathrm{ATP}}/\kt] \exp[\frac{1}{\kt}\int\limits_{\n}^{\n+\dn\thetap} -\frac{\partial V(\n,\x)}{\partial \n}\, \mathrm{d}n ] \nonumber \\ 
\fl &=\weq \exp [\Delta\mu_{\mathrm{ATP}}/\kt] \exp[(-\fk\dn^2\thetap^2/2 - \fk(\n-\x)\dn\thetap)/\kt] \label{wplus}
\end{eqnarray}
and
\begin{eqnarray} 
\fl\wminus(\n,\x) &= \weq \exp [(\Delta\mu_{\mathrm{ADP}}+\Delta\mu_{\PI})/\kt] \exp[\frac{1}{\kt}\int\limits_{\n}^{\n-\dn\thetam} -\frac{\partial V(\n,\x)}{\partial \n}\, \mathrm{d}n ] \nonumber \\
\fl &=  \weq \exp [(\Delta\mu_{\mathrm{ADP}}+\Delta\mu_{\PI})/\kt] \exp[(-\fk\dn^2\thetam^2/2 + \fk(\n-\x)\dn\thetam)/\kt]
\label{wminus}
\end{eqnarray}
with
\begin{equation}
 \Delta\mu_i\equiv\mu_i-\mu_i^{\mathrm{eq}}=k_BT\ln(c_i/c_i^{\mathrm{eq}})
\end{equation}
and $c_i$ the concentrations of the nucleotides ($i=$ ATP, ADP, $\PI$). Here, the transition rate $\weq$ applies to equilibrium concentrations of nucleotides. The load sharing factors $\thetap$ and $\thetam$, with $\thetap+\thetam=1$, depend on the unresolved shape of the free-energy landscape of the motor protein \cite{kolo07,fish99}.

\subsection{Fokker--Planck equation}
The transition rates, as well as the force the motor protein exerts on the probe, depend on the distance 
\begin{equation}
\y(t)\equiv\n(t)-\x(t) 
\end{equation}
between motor and probe. The corresponding probability density $p(y,t)$ obeys the Fokker--Planck--type equation
\begin{eqnarray}
\fl \partial_t p(\y,t) & = \wplus(\y-\dn)\, p(\y-\dn,t)+\wminus(\y+\dn)\, p(\y+\dn,t)-(\wplus(\y)+\wminus(\y))\, p(\y,t) \nonumber \\ \fl & \quad \quad +\partial_\y ((\fk\y-\fext)\, p(\y,t)+ \kt\partial_\y \, p(\y,t))/\gamma,
\label{FPmastery}
\end{eqnarray}
which contains in the first line the contributions from the transitions of the motor protein and in the second line drift and diffusion of the probe particle. 

For large $t$ and constant nucleotide concentrations, the system reaches a stationary state with time independent $p^s(\y)$ and constant mean velocity $\mean{\dot{n}}=\mean{\dot{x}}\equiv v$ with
\begin{equation}
\mean{\dot{x}}=(\fk\mean{y}-\fext)/\gamma 
\end{equation}
and
\begin{equation} 
\mean{\dot{n}}=d\mean{\wplus(y)-\wminus(y)}
\label{vmean}                                                                                           
\end{equation}
where $\mean{...}$ denotes the average using the stationary distribution $p^s(\y)$, which, however, cannot be determined analytically.

\section{Efficiencies}
\subsection{First law: single trajectory}
Following the concept of stochastic thermodynamics \cite{seki98,seif07} one can assign a first law on the level of a single trajectory. If the probe moves a small distance $\Delta x$, the first law becomes
\begin{equation}
 \Delta q_{\mathrm{P}}=(-\frac{\partial V}{\partial x}-\fext)\Delta x=(\fk y-\fext)\Delta x
\label{1lawP}
\end{equation}
where $\Delta q_{\mathrm{P}}$ is the heat dissipated by the probe, $\fext\Delta x$ is the work against the external force and ($\partial_{x} V) \Delta x$ the change of the internal energy of the spring due to the motion of only the probe. A jump of the motor protein gives rise to a first law in the form of \cite{seif11} 
\begin{eqnarray}
 0&=&\Delta V + \Delta E_{\mathrm{Sol}}+\Delta q_{\mathrm{M}} \nonumber \\
  &=&\Delta V - \dmu + T\Delta S_{\mathrm{Sol}}+\Delta q_{\mathrm{M}}
\label{1lawM}
\end{eqnarray}
without a contribution of the internal energy of the motor as its internal energy does not change in the one--state model. The change of the internal energy of the spring is given by
\begin{equation}
 \Delta V \equiv V(n\pm d,x)-V(n,x)
\end{equation}
where the sign depends on the the direction of the jump. Due to ATP turnover, the internal energy of the solution changes by $\Delta E_{\mathrm{Sol}}=-\dmu+T\Delta S_{\mathrm{Sol}}$, where $\Delta S_{\mathrm{Sol}}$ is the change of the entropy of the solution. The heat dissipated by the motor protein in this transition is denoted by $\Delta q_{\mathrm{M}}$.

\subsection{First law: ensemble average}
On average, the chemical energy gained from ATP consumption that involves changes of the entropy of the solvent will be dissipated as heat $Q$ in the environment and/or is delivered as work against the external force. In the stationary state, the internal energy of the spring is constant on average. Taking the average rates of (\ref{1lawP}) and (\ref{1lawM}) and summing the two contributions, this first--law condition can be expressed as
\begin{equation}
 \dot{\dmu}=\qdotk+\qdotm+T\dot{S}_{\mathrm{Sol}}+\fext v
\label{muq}
\end{equation}
where the dot denotes a rate and 
\begin{equation}
 \dot{\dmu}\equiv-\mean{\dot{F}_{\mathrm{Sol}}}=\dmu v/d
\end{equation}
is the rate of free energy consumption. The rate of dissipated heat $\dot{Q}=\qdotk+\qdotm$ has two contributions. First, the heat flow through the motor protein is given by 
\begin{equation}
 \qdotm\equiv\mean{\dot{q}_{\mathrm{M}}}=\dot{\dmu}-\dot{V}_n-T\dot{S}_{\mathrm{Sol}}, 
\label{heatflowM}
\end{equation}
representing the fact that while jumping, the motor protein uses free energy from the hydrolysis to load the spring which corresponds to a change of the internal energy of the spring $\dot{V}_n$ with
\begin{eqnarray}
\fl \dot{V}_n  &\equiv \int_{-\infty}^{\infty}p^s(y)[\wplus(y)(V(y+d)-V(y))+\wminus(y)(V(y-d)-V(y))] \, \mathrm{d}y \\
\fl &= \frac{\fk d^2}{2}\mean{\wplus(y)+\wminus(y)}+\fk d\mean{y(\wplus(y)-\wminus(y))}.
\end{eqnarray}
The energy thus stored in the spring is then dissipated by the probe whose heat flow is given by
\begin{equation}
\qdotk\equiv\mean{\dot{q}_{\mathrm{P}}}=\mean{(\fk\y-\fext)\nu(y)}
\label{heatflow}
\end{equation}
where 
\begin{equation}
 \nu(y)\equiv((\fk\y-\fext)+ \kt\partial_y \ln p^s(y))/\gamma.
\label{nuy}
\end{equation}
is the local mean velocity of the probe for a given $\y$ \cite{hata01,spec06} which corresponds to the current arising from the motion of only the probe in (\ref{FPmastery}).

\subsection{Three different efficiencies}

We will now focus on three different definitions of efficiency that have been proposed for motor proteins.

In the absence of an external force ($\fext=0$), one can compare the energy that the motor protein transfers to the spring, $\dot{V}_n$, with its available chemical energy $\dot{\dmu}$. From (\ref{muq}) and (\ref{heatflowM}) it follows that $\dot{V}_n=\qdotk$. The ratio of on average dissipated heat through the probe and available free energy 
\begin{equation}
 \eta_{\mathrm{Q}}\equiv\frac{\qdotk}{\dot{\dmu}}=\frac{d\fk\mean{y\nu}}{v\Delta\mu}
\end{equation}
was proposed as definition of efficiency \cite{toya10}. We will see below that $\etaq$ is not bounded by 1, as it has been anticipated earlier \cite{wang02a,kino04}, and therefore we will call it a pseudo efficiency. A second type of efficiency is the Stokes efficiency, 
\begin{equation}
 \eta_{\mathrm{S}}\equiv\frac{\gamma v^2}{\dot{\dmu}}=\frac{d\fk\mean{y}}{\Delta\mu},
\label{etastk}
\end{equation}
that compares the mean drag force $\gamma v$ the probe feels with the available chemical force. In contrast to $\etaq$, $\etastk$ is bounded by 1 \cite{wang02a}. If the motor protein exerted a constant force on the probe, the Stokes efficiency would be equal to the pseudo efficiency $\etaq$ because in this case the average heat dissipated by the probe is the mean drag force times $d$. 

Finally, in the presence of an external force acting on the probe, the thermodynamic efficiency of the system is the ratio between mechanical work delivered to the external force and available free energy \cite{parm99}
\begin{equation}
\eta_{\mathrm{T}}\equiv \frac{\fext v}{\dot{\dmu}} =\frac{\fext d}{\Delta\mu}.
\end{equation}
For $\fext\neq 0$, the pseudo efficiency $\etaq$ can be defined as 
\begin{equation}
\eta_{\mathrm{Q}}=\frac{\qdotk+\fext v}{\dot{\dmu}}.
\label{etaqextf}
\end{equation}

\section{Gaussian approximation}
\subsection{Derivation}

For a comparison with the simulations and in order to gain more analytical insights, it will be convenient to have a simple approximation for the stationary distribution $p^s(y)$. For a Gaussian probability distribution
\begin{equation}
 p^G(y)\equiv\frac{1}{\sqrt{2\pi}\sigma}\exp[-\frac{(y-\bar{y})^2}{2\sigma^2}]
\label{gauss}
\end{equation}
the free parameters $\bar{y}$ for the mean and $\sigma^2$ for the variance can be determined by requiring that the time--derivative of these quantities as calculated with the Fokker--Planck equation (\ref{FPmastery}) vanishes in the steady state. These conditions result in the following two equations for $\bar{y}$ and $\sigma^2$
\begin{equation}
(\fk\bar{y}-\fext)/\gamma=d(\wgplus-\wgminus)
\label{Gaussgly}
\end{equation}
and
\begin{eqnarray}
\fl (\fk\sigma^2+\fk\bar{y^2}-\fext\bar{y}-k_BT)/\gamma=&d[(\bar{y}-k d \thetap\sigma^2/\kt)\wgplus-(\bar{y}+k d \thetam\sigma^2/\kt)\wgminus]\nonumber \\\fl &+d^2(\wgplus+\wgminus)/2
\label{Gaussgls},
\end{eqnarray}
where we have introduced the average jump rates
\begin{eqnarray}
\fl \wgplus&\equiv \int_{-\infty}^{\infty}\wplus(y) p^G(y)\,\mathrm{d}y \nonumber \\ \fl &=\weq\exp[\Delta\mu_{\mathrm{ATP}}/\kt-\fk d^2\thetap^2(\kt-\fk\sigma^2)/2(\kt)^2-\fk d\thetap\bar{y}/\kt] \label{wgp}\\
\fl \wgminus&\equiv \int_{-\infty}^{\infty}\wminus(y) p^G(y)\,\mathrm{d}y \nonumber \\ \fl &=\weq\exp[(\Delta\mu_{\mathrm{ADP}}+\Delta\mu_{\PI})/\kt-\fk d^2 \thetam^2(\kt-\fk\sigma^2)/2(\kt)^2+\fk d \thetam\bar{y}/\kt] \label{wgm}.
\end{eqnarray}
These equations can easily be solved numerically.

\subsection{Limits $\dmu\rightarrow 0$ and $\dmu\rightarrow\infty$}
Close to chemical equilibrium, i.e., $\Delta\mu=0$, and for $\fext=0$, we expand $\bar{y}$ and $\fk\sigma^2-\kt$ up to first order in $\Delta\mu$ and find
\begin{equation}
 \bar{y}\approx A\Delta\mu+\tilde{A}(\thetap-\thetam)^2\Delta\mu
\label{mu0y}
\end{equation}
and
\begin{equation}
 \fk\sigma^2-\kt\approx B(\thetap-\thetam)\Delta\mu.
\label{mu0ks}
\end{equation}
The coefficients $A$, $\tilde{A}$ and $B$ obtained by solving the first order of (\ref{Gaussgly}) and (\ref{Gaussgls}) are too long to be shown here.

In the limit $\dmu\rightarrow\infty$ and $\fext=0$, we obtain for $\bar{y}$ and $\fk\sigma^2-\kt$ 
\begin{equation}
 \bar{y}\approx \frac{\dmu}{\fk d}-C\dmu\exp[-\dmu\thetam/\kt]
\end{equation}
and
\begin{equation}
 \fk\sigma^2-\kt\approx D\dmu\exp[-\dmu\thetam/\kt]
\label{muinfy}
\end{equation}
as long as $\thetam>0$. The coefficients 
\begin{equation}
 C=\frac{1+\fk d^2(\thetap-\thetam)^2/4}{\gamma \fk d^3\weq\exp[(\dmu_{\mathrm{ADP}}+\dmu_{\PI})/\kt]}
\label{muinfks}
\end{equation}
and
\begin{equation}
 D=-\frac{\thetap-\thetam}{2\gamma d^2\weq\exp[(\dmu_{\mathrm{ADP}}+\dmu_{\PI})/\kt]}
\end{equation}
are obtained by solving (\ref{Gaussgly}) and (\ref{Gaussgls}) to first and second order in $\dmu$.

\subsection{Efficiencies}

Within this Gaussian approximation, the average heat flow through the probe as given by (\ref{heatflow}) is calculated using the local mean velocity (\ref{nuy})
\begin{equation}
 \nu(y)=(\fk\y-\fext)/\gamma-\kt(y-\bar{y})/(\gamma\sigma^2).
\end{equation}
The average over $y$ can now be performed leading to 
\begin{equation}
\qdotk =(\fk^2\sigma^2+\fk^2\bar{y}^2-\kt\fk-2\fk\fext\bar{y}+\fext^2)/\gamma. 
\end{equation}
This expression is used to determine $\etaq$ in this approximation as
\begin{equation}
 \etaq= d\fk\frac{\fk\sigma^2-\kt+\fk\bar{y}^2-\fext\bar{y}}{\Delta\mu(\fk\bar{y}-\fext)}
\label{etaqG}
\end{equation}
with $\bar{y}$ and $\sigma^2$ being the solution of (\ref{Gaussgly}) and (\ref{Gaussgls}) for given $\Delta\mu$ and $\weq$.

For small $\dmu$, using (\ref{mu0y}) and (\ref{mu0ks}), $\etaq$ takes the form
\begin{equation}
 \etaq\approx\frac{dB(\thetap-\thetam)}{(A+\tilde{A}(\thetap-\thetam)^2)\dmu}+\fk dA+\fk d\tilde{A}(\thetap-\thetam)^2.
\end{equation}
If $\thetap\neq\thetam$, $\etaq$ diverges for vanishing $\dmu$. For $\thetap>\thetam$, $\etaq$ can become negative due to those jumps of the motor protein that occur when the previous diffusion of the probe has resulted in $y<-0.5d$. Then, the energy stored in the spring is dissipated by the motor protein during jumping.

In the limit of large $\dmu$, we use (\ref{muinfy}) and (\ref{muinfks}) to obtain
\begin{equation}
 \etaq\approx 1-\fk dC\exp[-\dmu\thetam/\kt]+\frac{dD\exp[-\dmu\thetam/\kt]}{\dmu/(\fk d)-C\dmu\exp[-\dmu\thetam/\kt]}
\end{equation}
which approaches 1.

The Stokes efficiency in the Gaussian approximation without an external force is simply given by
\begin{equation}
 \etastk=\frac{d\fk\bar{y}}{\Delta\mu}.
\end{equation}
For $\fk\sigma^2>\kt$, which is the case for $\thetap<0.5$, the Stokes efficiency is always smaller than $\etaq$. For vanishing $\dmu$, $\etastk$ approaches a finite value, $\etastk\approx d\fk A+d\fk \tilde{A}(\thetap-\thetam)^2$, while for $\dmu\rightarrow\infty$ it also converges to 1.

\section{Results}
\label{Results}

In this section, we study the three efficiencies for our hybrid model as functions of the chemical energy $\dmu$, the absolute concentrations of the nucleodides, i.e. $\weq$, the external force $\fext$ and the load sharing factor $\thetap$. The data are obtained from simulations, using a Gillespie algorithm \cite{gill92} similar to \cite{gasp07} with the motion of the probe being spatially discretized in steps of $\Delta\x=\dn/1000$, and compared with the Gaussian approximation.

We use model parameters as given in table \ref{Para} which are motivated by experimental results for the $\feins$--ATPase as described in section \ref{Comparison} below. The load sharing factor $\thetap$ remains as a free parameter.

\begin{figure}[t]
 \centering
\includegraphics[height=8cm]{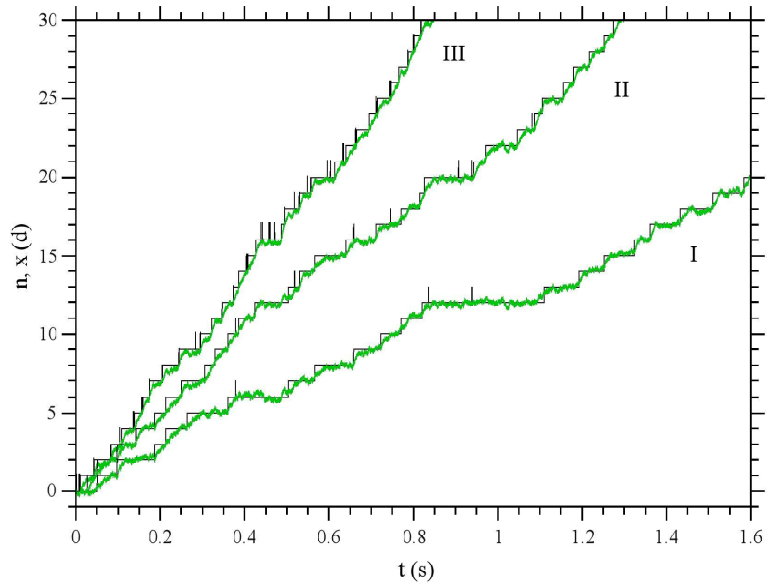}
\caption{Trajectories of the molecular motor (black) and the attached probe particle (green) obtained from the simualtion (without external force). In the presence of low nucleotide concentrations (I) the trajectory of the probe exhibits a more stepwise form while it becomes almost linear for high nucleotide concentrations (III). The parameters are $\thetap=0.1$, (I): $c_{\mathrm{ATP}}=0.4\,\mu\mathrm{M}$, $c_{\mathrm{ADP}}=0.4\,\mu\mathrm{M}$, $c_{\PI}=1$ mM, (II): $c_{\mathrm{ATP}}=2\,\mu\mathrm{M}$, $c_{\mathrm{ADP}}=2\,\mu\mathrm{M}$, $c_{\PI}=1$ mM, (III): $c_{\mathrm{ATP}}=100\,\mu\mathrm{M}$, $c_{\mathrm{ADP}}=100\,\mu\mathrm{M}$, $c_{\PI}=1$ mM. For all parameter sets we have $\dmu=19.14\,\kt$. The nucleotide concentrations are the same as used in the experiment \cite{toya10} shown in figure \ref{QWplot} with the same labelling (I-III) below.}
\label{traj}
\end{figure}

Simulated trajectories with the same nucleotide concentrations as used in the experiment \cite{toya10} are shown in figure \ref{traj}. In the presence of low nucleotide concentrations only few backward jumps of the motor protein take place and the trajectory of the probe shows an almost staircase like form. For high nucleotide concentrations, following a forward step the motor often performs a backward jump. Such a sequence of two jumps is not necessarily visible in the trajectory of the probe which remains almost linear.

\Table{\label{Para}Values of the model parameters used for the simulation and the Gaussian approximation.}
\br
$\gamma$ ($\kt\mathrm{s}/d^2$) &  $\fk$ ($\kt/d^2$) & $\weq/c_{\mathrm{ATP}}^{\mathrm{eq}}$ ($\mathrm{M}^{-1}\mathrm{s}^{-1}$) \\
\mr 

$0.407$ & $40$ & $3\cdot 10^{7} $ \\

\br

\endTable

\subsection{Pseudo efficiency $\etaq$}

\begin{figure}[top]
 \centering
\subfigure[]{\includegraphics[height=5cm]{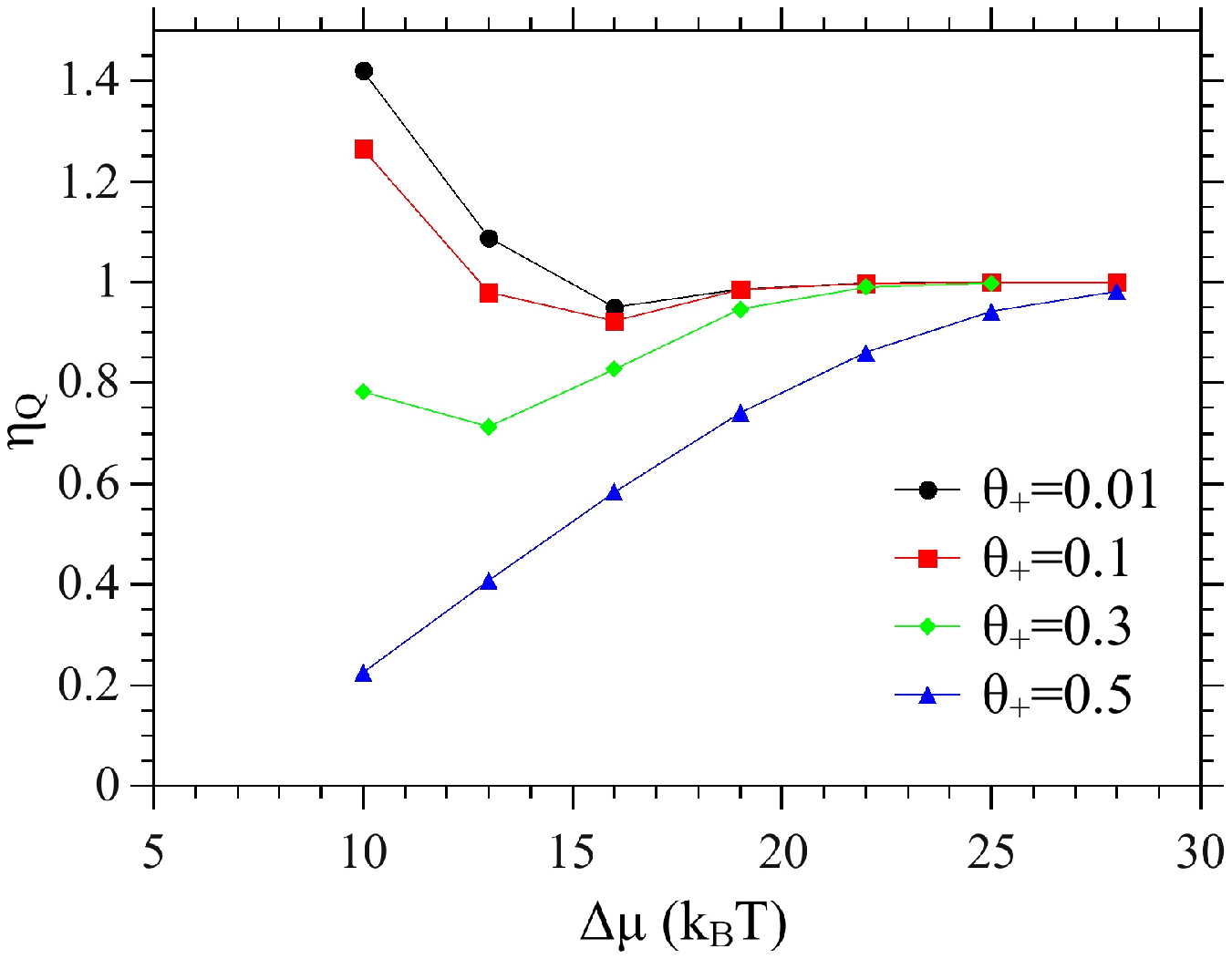}}
\subfigure[]{\includegraphics[height=5cm]{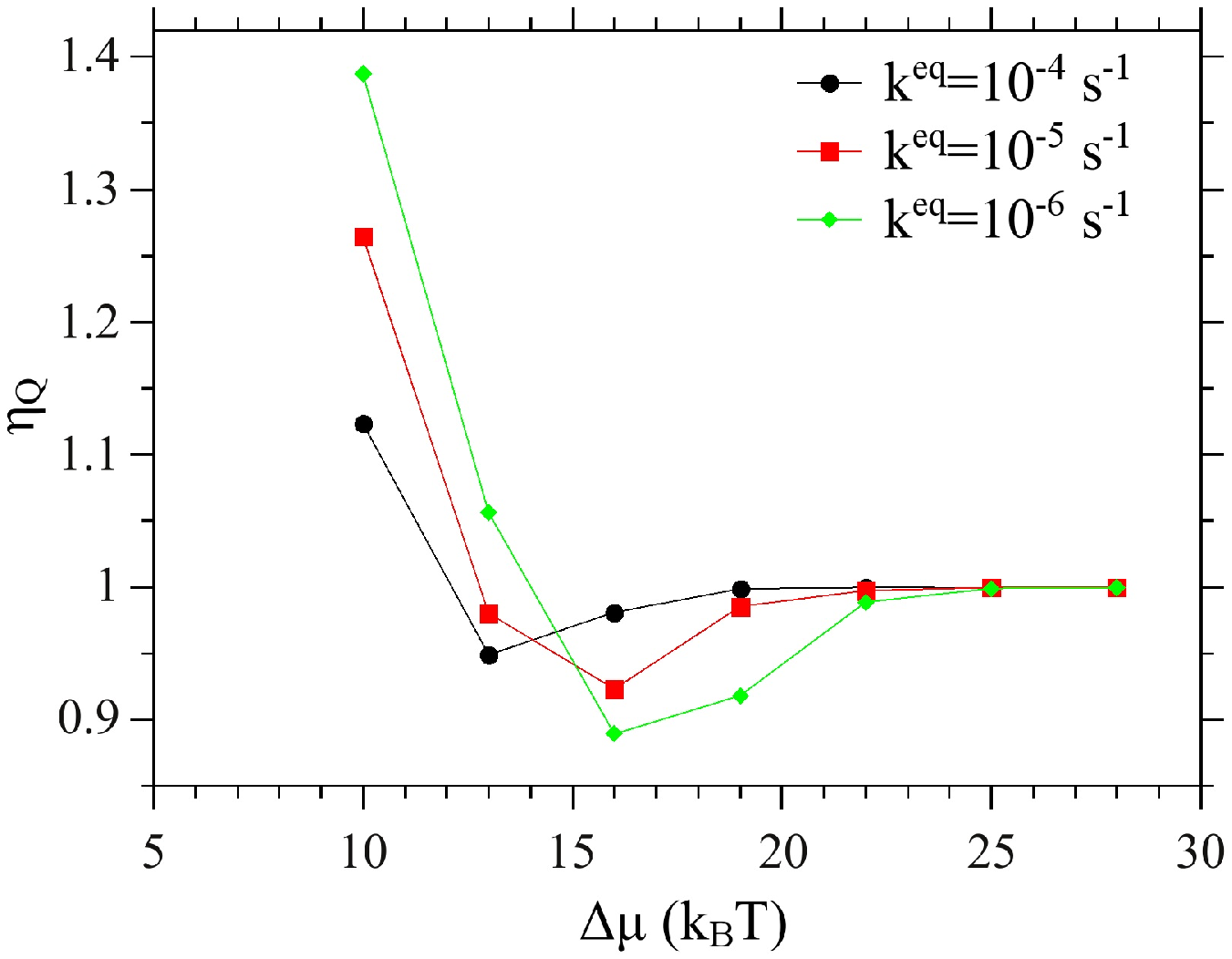}}
\subfigure[]{\includegraphics[height=5cm]{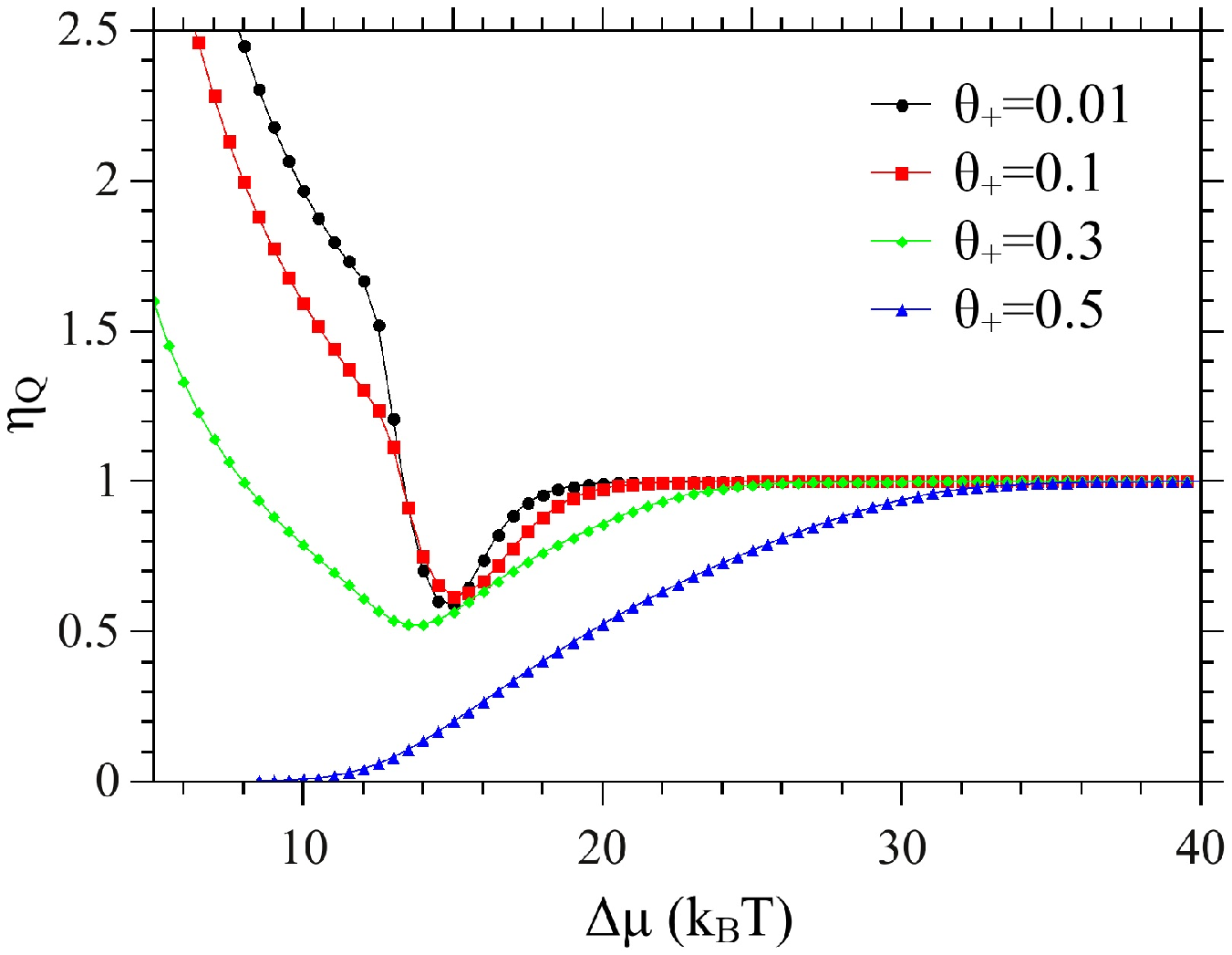}}
\subfigure[]{\includegraphics[height=5cm]{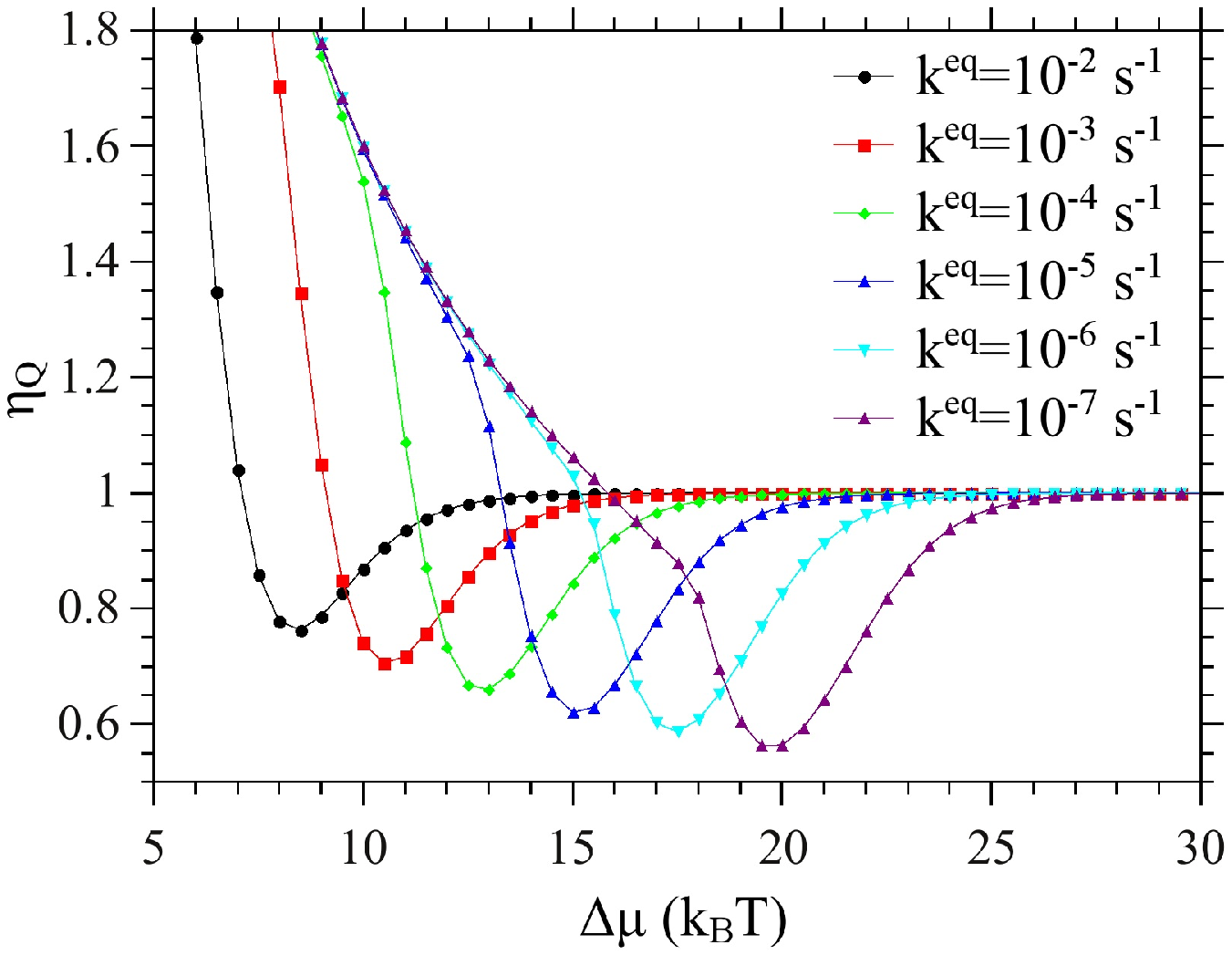}}
\caption{Pseudo efficiency $\eta_{\mathrm{Q}}$ from the simulation, (a) and (b); and within the Gaussian approximation, (c) and (d). (a) and (c): $\etaq$ as a function of $\Delta\mu$ for different values of the load sharing factor $\thetap$ and fixed $\weq=10^{-5}\mathrm{s}^{-1}$. (b) and (d): $\eta_{\mathrm{Q}}$ as a function of $\Delta\mu$ for different values of $\weq$ with fixed $\thetap=0.1$. The remaining parameters are $\fk=40\kt/d^2$, $\gamma=0.407\kt\mathrm{s}/d^2$. In the simulation, the error is of the order of the symbol size.}
\label{Etaq}
\end{figure}

\begin{figure}[top]
 \centering
\includegraphics[height=8cm]{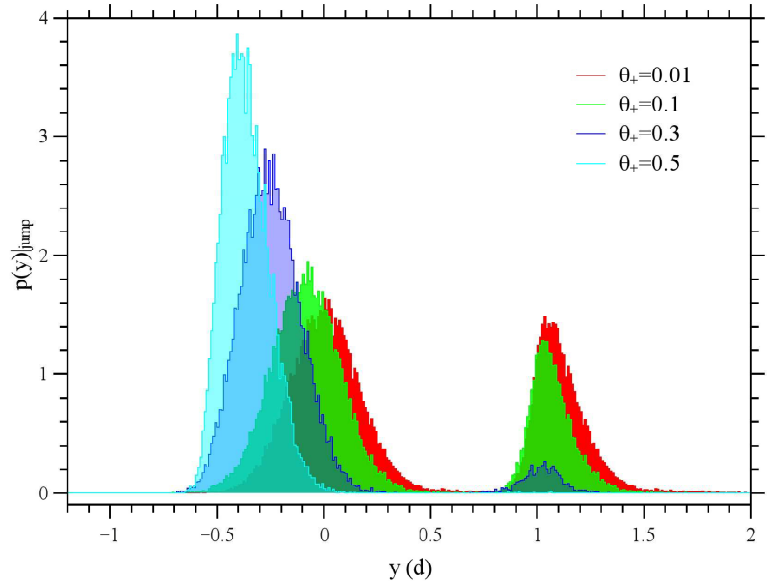}
\caption{Probability distribution $p(y)_{|\mathrm{jump}}$ of the distance $y$ just before a jump of the motor protein for several values of $\thetap$ at $\Delta\mu=13\kt$ and $\weq=10^{-5}\mathrm{s}^{-1}$. For small $\thetap$, the forward jumps of the motor protein are almost independent of the position of the probe resulting in a peak at $y\simeq0$ whereas for larger $\thetap$ the peak clearly shifts to $y<0$ implying that the motor protein prefers to jump when the probe has diffused ahead. The peaks around $y=1$ indicate backward jumps which take place more often in the case of small $\thetap$ when the backward rate is more sensitive to the position of the probe.}
\label{Hyjumps}
\end{figure}

We will first investigate the pseudo efficiency $\eta_{Q}$ as a function of $\dmu$, $\weq$ and $\thetap$. We extract $\qdotk$ from the numerical data by averaging over one sufficiently long trajectory. The results are shown in figure \ref{Etaq}. The most striking fact of these data is the observation that $\etaq$ is larger than 1 for small enough $\dmu$ and $\thetap$ which shows up in the Gaussian approximation as well. This effect can be understood as follows. In a jump, the motor protein can take heat from the solution in order to change the internal energy of the spring by an amount larger than $\dmu$. If, subsequently, the probe dissipates this internal energy of the spring as heat back into the environment, $\qdotk$ can indeed become larger than $\dot{\dmu}$ without any violation of the second law. Using the obtained parameter for the spring constant $\fk$, the motor protein transfers $20\,\kt$ to the spring if it starts the jump from the minimum of the harmonic potential. For small values of $\thetap$, the forward jump rate of the motor protein depends only weakly on the current position of the probe as shown in figure \ref{Hyjumps}. Therefore, jumps will occur even if the associated change of internal energy of the spring, $\Delta V$, is larger than $\Delta\mu$. For rather small $\weq$, backward jumps are rare and the probe relaxes to the potential minimum between successive forward jumps (see data set I in figure \ref{traj}). This leads to $\dot{V}_n>\dot{\dmu}$ on average and hence to $\eta_{Q}>1$ for $\Delta\mu$ considerably smaller than $20\,\kt$ as shown in figure \ref{Etaq}. As the value of $\thetap$ increases, $\etaq$ decreases because the forward jumps of the motor protein are suppressed. On average, in this case the motor protein jumps only if the probe has diffused forward and exerts a pulling force on the motor through the spring.

\begin{figure}[top]
 \centering
\subfigure[]{\includegraphics[height=5cm]{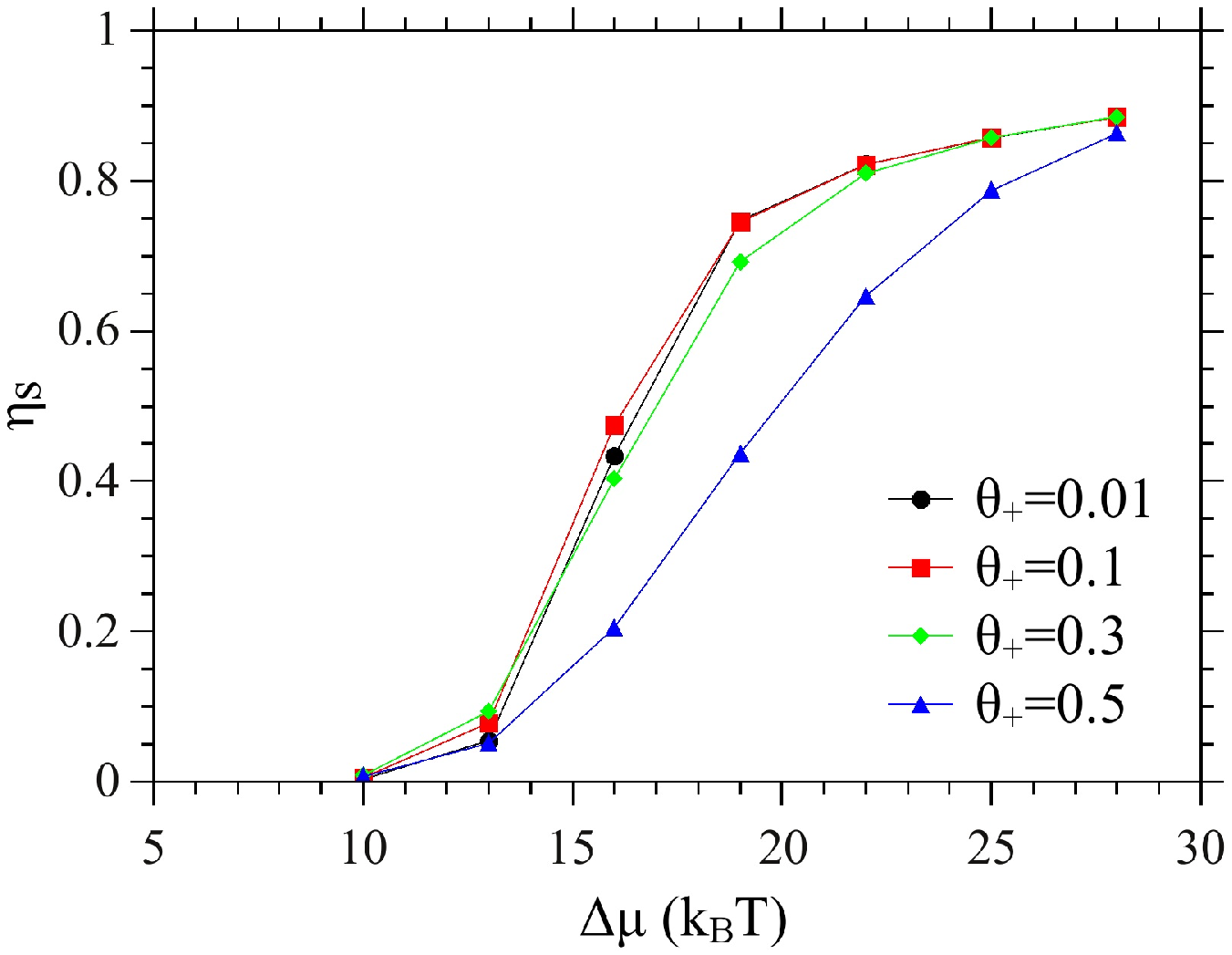}} 
\subfigure[]{\includegraphics[height=5cm]{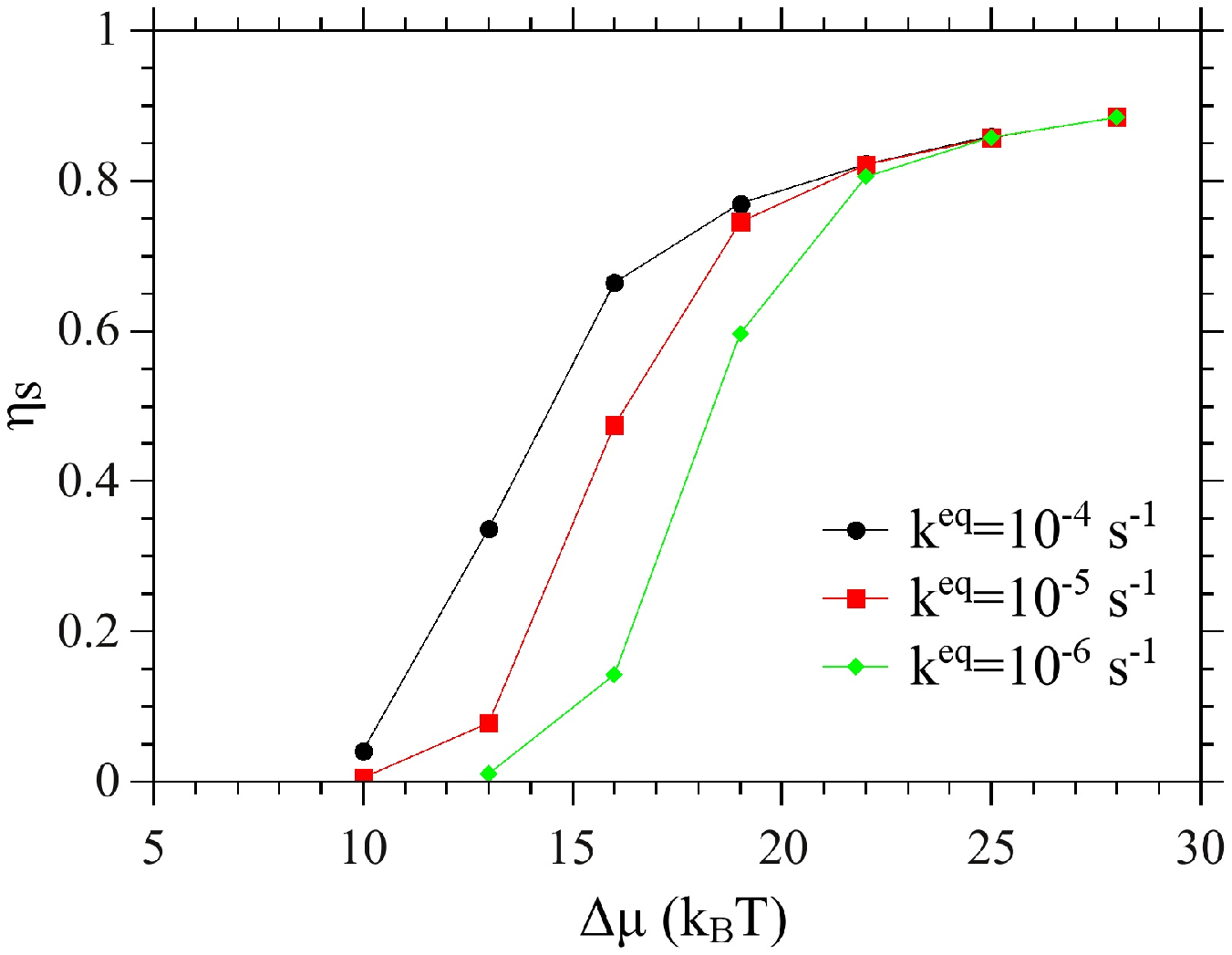}}
\subfigure[]{\includegraphics[height=5cm]{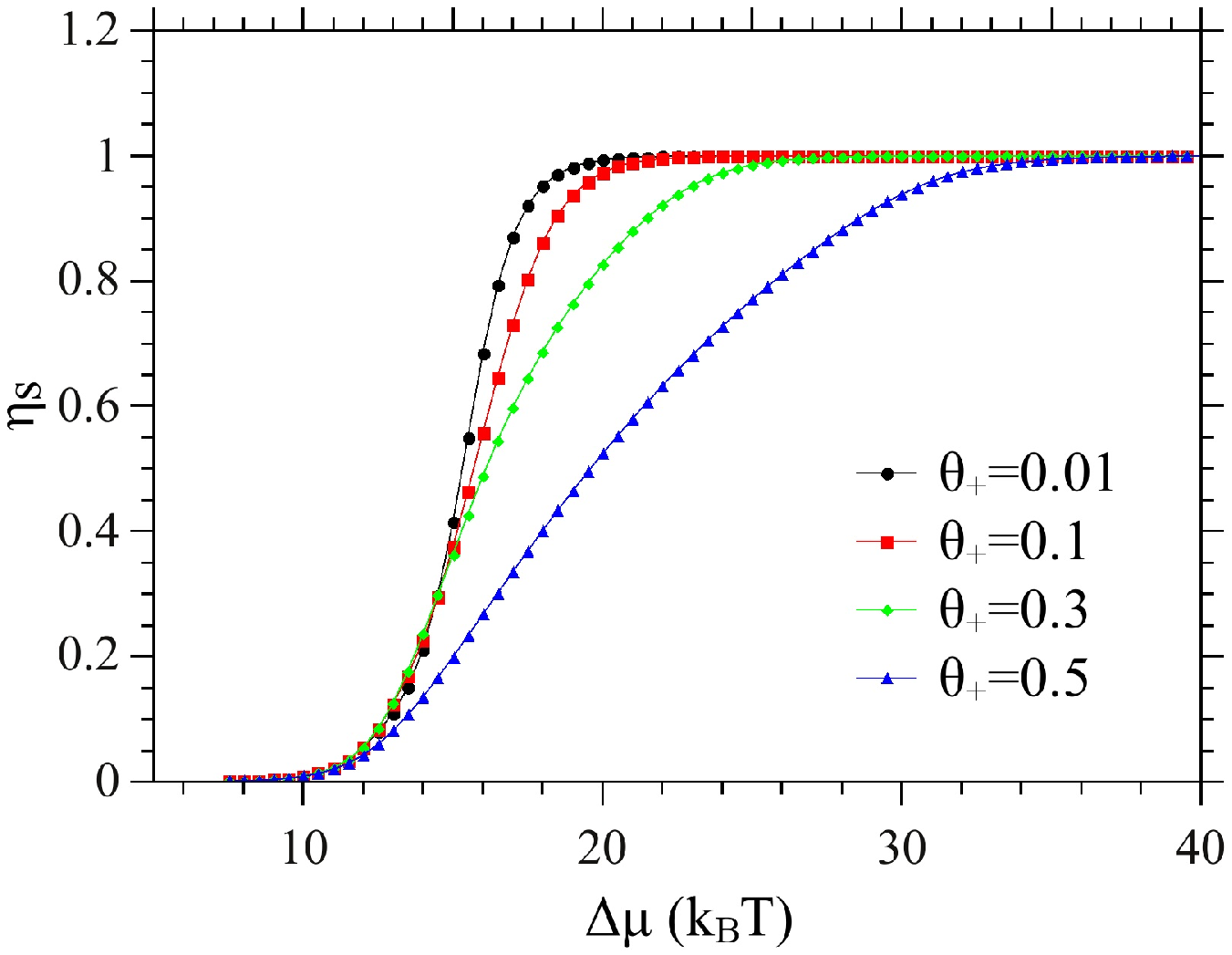}}
\subfigure[]{\includegraphics[height=5cm]{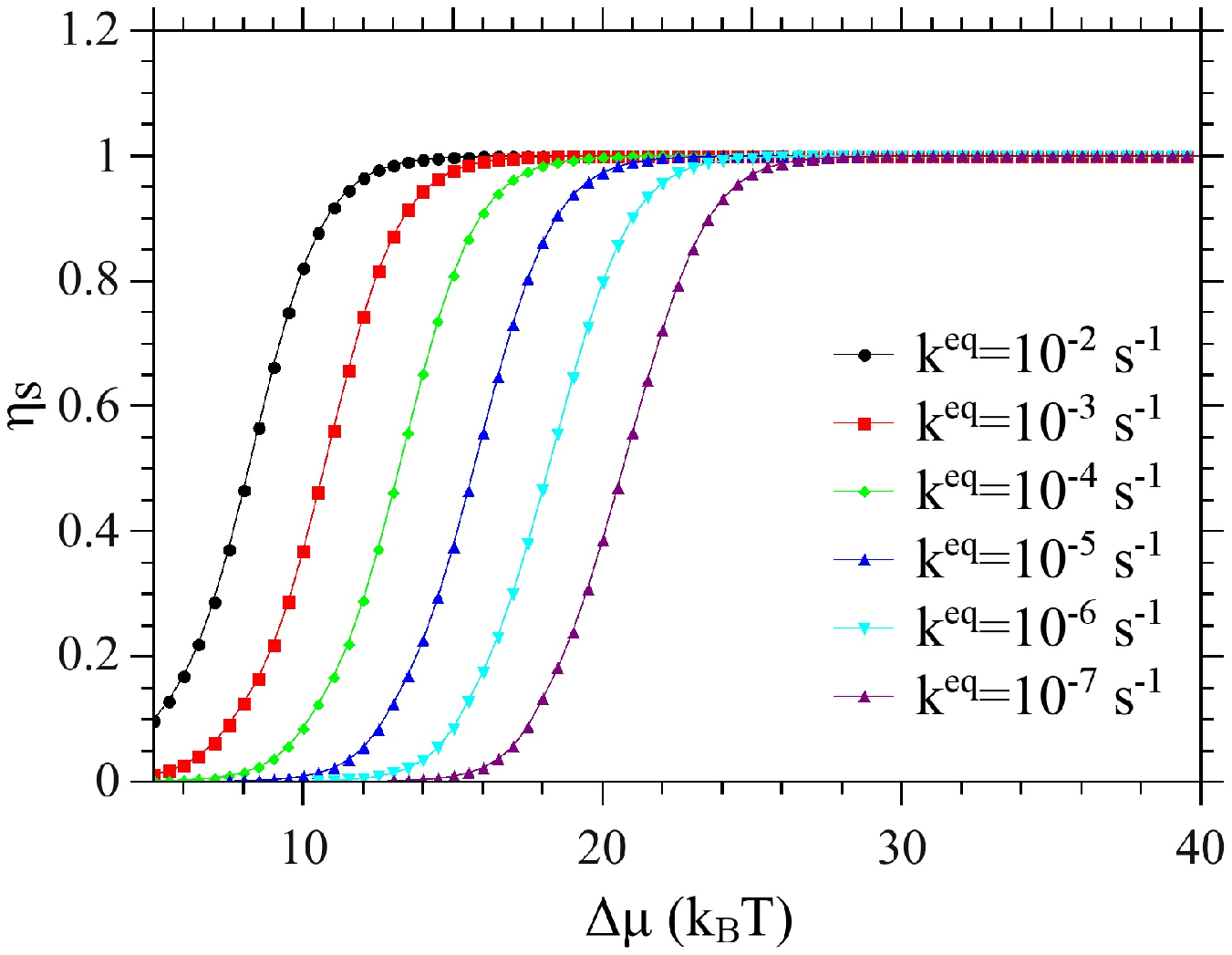}}
\caption{(a) Stokes efficiency $\etastk$ from the simulation, (a) and (b); and within the Gaussian approximation, (c) and (d). The data is obtained from the same trajectories used to obtain $\etaq$ in figure \ref{Etaq} (a) and (b). (a) and (c): $\etastk$ as a function of $\Delta\mu$ for different values of the load sharing factor $\thetap$ and fixed $\weq=10^{-5}\mathrm{s}^{-1}$. (b) and (d): $\etastk$ as a function of $\Delta\mu$ for different values of $\weq$ for fixed $\thetap=0.1$.}
\label{Etastk}
\end{figure}

Increasing the absolute concentrations of the nucleotides, i.e., increasing $\weq$, results in more forward but also more backward jumps, which can be seen for data set III in figure \ref{traj}. For small $\dmu$, the occasional backward jumps follow especially those forward jumps for which the change of internal energy of the spring has been larger than $\Delta\mu$, leading to a smaller $\etaq$.

In the limit of large $\dmu$, the motor protein jumps even when the spring is previously stretched which can result in changes of the internal energy of the spring by an amount larger than $20\,\kt$. The coupling between the motor protein and the probe induces a balancing effect between the forward motion of the motor protein and the drag of the probe maintaining a typical $\dot{V}$ that turns out to be approximately $\dot{\dmu}$, leading to $\etaq\simeq 1$.

\subsection{Stokes efficiency $\etastk$}
We also obtain the Stokes efficiency (\ref{etastk}) from the simulated trajectories and the Gaussian approximation as shown in figure \ref{Etastk}. Characteristically, starting close to 0 for small $\dmu$, $\etastk$ monotonically increases with $\dmu$ reaching 1 for $\dmu\rightarrow\infty$. For small $\Delta\mu$, the trajectory of the probe shows a staircase form with small average velocity leading to small values of the Stokes efficiency in contrast to values of the pseudo efficiency $\etaq>1$. For large $\Delta\mu$, the probe does not relax to the potential minimum between consecutive jumps resulting in a more linear trajectory of the probe as if it was exposed to an almost constant force. In this limit of an almost linear motion of the probe, the pseudo efficiency becomes the Stokes efficiency. As $\etastk$ is bounded by 1, $\etaq$ can not reach values larger than 1 in this limit either.

Increasing the load sharing factor $\thetap$ results in decreasing average velocities. Therefore, the Stokes efficiency also decreases which can be seen in figures \ref{Etastk} (a) and (c). With increasing absolute concentrations of nucleotides, i.e., with increasing $\weq$, the average velocity and therefore also the Stokes efficiency at fixed $\Delta\mu$ increases as shown in figures \ref{Etastk} (b) and (d).

\subsection{Thermodynamic efficiency $\eta_{\mathrm{T}}$}
\begin{figure}[top]
 \centering
\includegraphics[height=8cm]{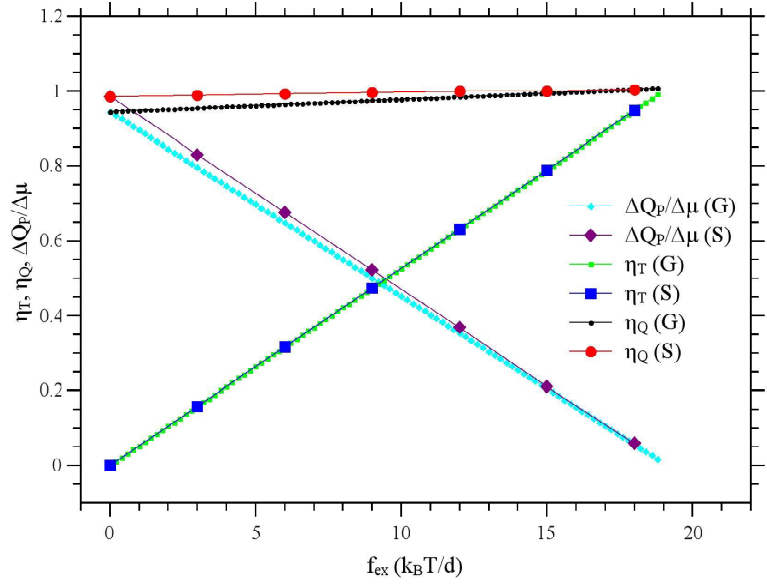}
\caption{Pseudo efficiency $\etaq$ (red and black dots), thermodynamic efficiency $\eta_{\mathrm{T}}$ (green and blue squares) and dissipated heat $\Delta \qk$ per $\dmu$ (cyan and purple diamonds) as functions of $\fext$ for fixed $\dmu=19\kt$, $\thetap=0.1$ and $\weq=10^{-5}\mathrm{s}^{-1}$. The plot contains data from the simulation (S) as well as from the Gaussian approximation (G).}
\label{etavonfex}
\end{figure}

The thermodynamic efficiency of the system can be studied only if an external force is applied to the probe. As an illustrative example, for fixed $\dmu=19 \kt$, we examine the thermodynamic efficiency in the presence of external forces smaller than the stall force as shown in figure \ref{etavonfex}. The thermodynamic efficiency increases linearly with $\fext$ and reaches 1 at the stall force which is possible only due to the tight mechanochemical coupling in this model. In figure \ref{etavonfex} we also show the pseudo efficiency $\etaq$ as defined in (\ref{etaqextf}) in the presence of external forces. While $\etaq$ is almost independent of the external force, the contribution from the dissipated heat, $\qdotk/\dot{\dmu}$, decreases linearly with $\fext$ and reaches zero at the stall force.

At stall conditions, the work corresponding to the stall force refers to the maximum work the motor protein can convert on average. In the simulation, we find that applying $\fext=\dmu/d$ generates a diffusive motion of the motor protein with $v\simeq 0$ and $\qdotk\simeq 0$ for various values of $\thetap$ and $\dmu$, including the ones with $\etaq>1$ and $\etaq<1$. This observation implies that the motor protein seems in principle to be able to convert the full $\dmu$ into extractable work. In our model, where the motor interacts with the external force only via the spring, this result is not trivial, as it would have been if we had applied $\fext$ directly to the motor. Under these conditions, $p^s(y)$ is Gaussian with $\mean{y}=\dmu/(dk)$ and the same variance as the Boltzmann-distribution in equilibrium, $\sigma^2=\kt/\fk$.

Within the Gaussian approximation we can insert $\fext=\dmu/d$ in (\ref{Gaussgly}) and (\ref{Gaussgls}). With $\fk\sigma^2=\kt$, $\bar{y}=\Delta\mu/(d\fk)$, i.e., $v=0$ is a solution for $\fext=\dmu/d$, implying that also in the Gaussian approximation the motor protein is able to convert the full $\Delta\mu$ into extractable work for any values of the load sharing factors given that $\thetap+\thetam=1$.

\section{Case study: $\feins$--ATPase}
\label{Comparison}

In this section, we apply our hybrid model to the $\feins$--ATPase and compare the simulations with recent experimental data \cite{toya10,toya11,toya11a}.

\subsection{Model parameters}
\label{model_param}

\begin{figure}[top]
 \centering
\includegraphics[height=8cm]{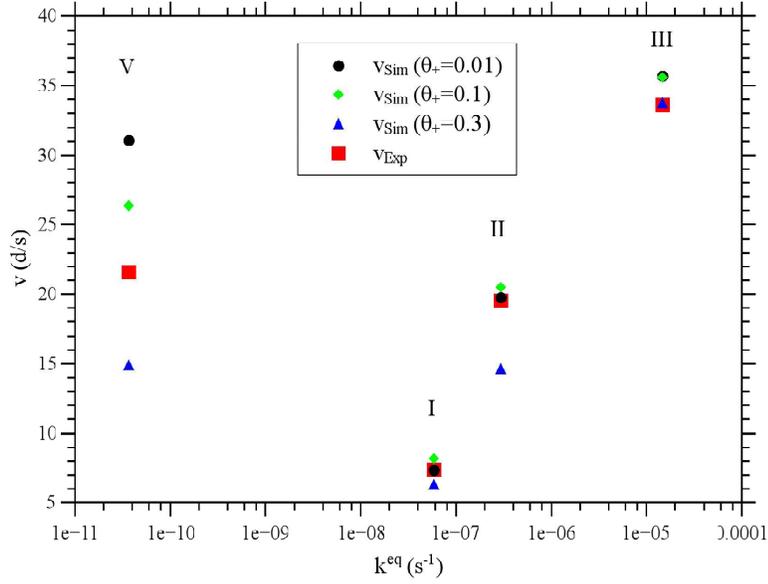}
\caption{Comparison of the mean velocities observed experimentally \cite{toya10} (red squares) and in the simulation for several load sharing factors $\thetap$ (black dots, green diamonds and blue triangles). The labelling I, II, III, V refers to the corresponding parameter sets in figure \ref{QWplot}.}
\label{vmeanplot}
\end{figure}

For a quantitative comparison we have to map the rotary motion of the $\feins$--ATPase to our linear model and determine the model parameters. In our model, one jump of the motor protein covering a distance $d$ corresponds to a rotation of the $\gamma$ shaft of $120^{\circ}$. Using large probe particles like polystyrene beads or actin filaments, the substeps in one $120^{\circ}$ rotation are not resolved experimentally. Therefore, we will omit the substeps here, too. We assume that the temperature of the solution is $T\simeq24^{\circ}$C and that the probe consists of two beads of diameter $287$nm \cite{toya10}. The friction coefficient of the probe can be calculated using the formula for the rotational frictional coefficient $\Gamma$ from \cite{haya10,gasp07} with the viscosity of water ($\eta\simeq 0.001 \mathrm{Ns}/\mathrm{m}^2$). The frictional torque $N=\Gamma\dot{\varphi}$ acting on the probe with angular velocity $\dot{\varphi}$ corresponds to a frictional force 
\begin{equation}
 f_{\mathrm{fr}}=\frac{\Gamma}{r^2}\dot{x}=\gamma\dot{x}
\end{equation}
acting on the probe at distance $r$ from the $\gamma$ shaft. Within one $120^{\circ}$ rotation, the probe at distance $r$ covers $d=2\pi r /3$. For the linear model, the friction coefficient $\gamma$ can be calculated as 
\begin{equation}
 \gamma=\frac{\Gamma}{r^2}=\frac{4\pi^2\Gamma}{9d^2}
\end{equation}
leading to $\gamma=0.407\,\kt\,\mathrm{s}/d^2$. 

Following the mass action law assumption, the equilibrium transition rate $\weq$ is supposed to depend linearly on the concentrations of nucleotides in the solvent. For low ATP concentrations ($c_\mathrm{ATP}\simeq 10^{-6}$M), the mean velocity of the motor protein is dominated by the rate of ATP binding. In the one--step model this feature holds for all concentrations. Therefore we choose $\weq$ to be the experimentally determined rate of ATP binding $\weq\simeq3\cdot 10^7 \mathrm{M}^{-1}\mathrm{s}^{-1} c_{\mathrm{ATP}}^{\mathrm{eq}}$\cite{yasu01}. For known nonequilibrium concentrations of nucleotides like in the experiments, the structure of the transition rates (\ref{wplus}) and (\ref{wminus}) leaves the choice of the equilibrium concentrations arbitrary as long as they obey
\begin{equation}
 \frac{c_{\rm {ATP}}^{\mathrm{eq}}}{c^{\mathrm{eq}}_{\rm {ADP}}c^{\mathrm{eq}}_{\rm P_{\mathrm i}}}\simeq 4.89\cdot 10^{-6}\frac{1}{\mathrm{M}}
\end{equation}
for $T=23^\circ \mathrm{C}$ and pH $7$ \cite{gasp07}. For given $\weq$ and $\dmu$, one possible choice of the nonequilibrium concentrations of nucleotides is $c_{\mathrm {ADP}}=c^{\mathrm{eq}}_{\mathrm {ADP}}$, $c_{\mathrm P_{\mathrm i}}=c^{\mathrm{eq}}_{\mathrm P_{\mathrm i}}$ and $c_{\mathrm{ATP}}=c^{\mathrm{eq}}_{\mathrm{ATP}}\exp[\dmu/\kt]$ which was used for the simulation and the Gaussian approximation.

In order to determine the spring constant $\fk$ and the load sharing factor $\thetap$ we use both the experimental data of the mean velocities \cite{toya10} and the histogram of the angular position of the probe at a jump \cite{toya11a}. While both data sets depend on both parameters, the velocity, especially for large $\weq$, is more sensitive to $\fk$ whereas the peak position of the histogram mainly depends on $\thetap$. Therefore, we primarily use the velocity data to fit $\fk$ and determine the load sharing factor $\thetap$ by comparing the peak position of the experimental histogram \cite{toya11a} with the left peak position of the corresponding histograms obtained by our simulation as the ones shown in figure \ref{Hyjumps}.

As a result, we obtain $\fk=40\pm5\,\kt/d^2$ and a value of $\thetap$ in the range $0.1\lesssim\thetap\lesssim0.3$. In figure \ref{vmeanplot}, we show how for this value of $\fk$ changing the load sharing factor affects the mean velocity for which we get the best overall agreement for $\thetap=0.1$. For later purposes, we also include data for $\thetap=0.01$.

\subsection{Comparison of efficiencies with experimental data}

\begin{figure}[top]
 \centering

\includegraphics[height=8cm]{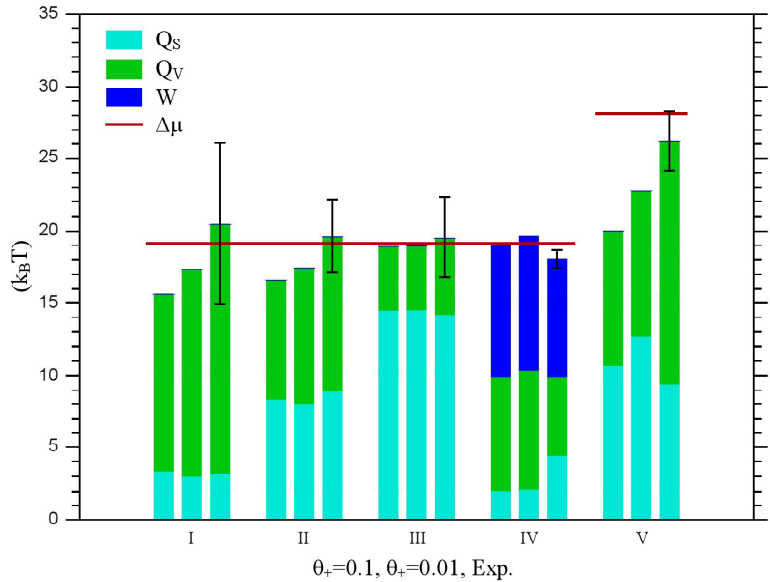}
\caption{Average heat $\qk$ released through the probe (green and cyan) and work $W\equiv\fext d$ against the external force (blue) compared to the available free energy per step $\dmu$ (red line). The dissipated heat of the probe is split into two contributions $Q_{\mathrm{S}}$ and $Q_{\mathrm{V}}$ according to the two terms of the Harada--Sasa relation (\ref{QSQV}). The contribution from the linear motion with constant mean velocity, $Q_{\mathrm{S}}$ (cyan), appears in the numerator of the Stokes efficiency while $Q_{\mathrm{V}}$ (green) is the contribution due to the non--uniform jumping motion of the motor protein. In each of the five parameter sets labelled by I-V, the left and the central bar represent results from the simulation for $\thetap=0.1$ and $\thetap=0.01$, respectively, while the right bar shows the experimental results and error bars from \cite{toya10}. The following parameters were used in the five cases: (I) $c_{\mathrm{ATP}}=0.4\,\mu\mathrm{M}$, $c_{\mathrm{ADP}}=0.4\,\mu\mathrm{M}$, $c_{\PI}=1$ mM, i.e, $\weq= 5.87\cdot10^{-8}\,\mathrm{s}^{-1}$ and $\dmu=19.14\,\kt$; (II)  $c_{\mathrm{ATP}}=2\,\mu\mathrm{M}$, $c_{\mathrm{ADP}}=2\,\mu\mathrm{M}$, $c_{\PI}=1$ mM, i.e, $\weq= 2.93\cdot10^{-7}\,\mathrm{s}^{-1}$ and $\dmu=19.14\,\kt$; (III) $c_{\mathrm{ATP}}=100\,\mu\mathrm{M}$, $c_{\mathrm{ADP}}=100\,\mu\mathrm{M}$, $c_{\PI}=1$ mM, i.e, $\weq= 1.47\cdot10^{-5}\,\mathrm{s}^{-1}$ and $\dmu=19.14\,\kt$; (IV) $c_{\mathrm{ATP}}=2\,\mu\mathrm{M}$, $c_{\mathrm{ADP}}=2\,\mu\mathrm{M}$, $c_{\PI}=1$ mM, $\fext=9.27\,k_BT/d$, i.e, $\weq= 2.93\cdot10^{-7}\,\mathrm{s}^{-1}$ and $\dmu=19.14\,\kt$; (V) $c_{\mathrm{ATP}}=2\,\mu\mathrm{M}$, $c_{\mathrm{ADP}}=0.5\,\mu\mathrm{M}$, $c_{\PI}=0.5\,\mu\mathrm{M}$, i.e, $\weq= 3.67\cdot10^{-11}\,\mathrm{s}^{-1}$ and $\dmu=28.12\,\kt$;}
\label{QWplot}
\end{figure}

Experimentally, the heat flow of the probe is determined using the Harada--Sasa relation \cite{hara05}. In the appendix, we show that this heat flow is equal to $\qdotk$ as defined in (\ref{heatflow}). In figure \ref{QWplot}, we plot the average heat released through the probe per step, $\qk$, plus the work against the external force, $W$, obtained by the simulation for $\fk=40\,\kt/d^2$ and $\thetap=0.1$ and compare it with the experimental results \cite{toya10}. We find quite good agreement between theory and experiment for the parameter sets I-IV where either the maximum deviation is $15\%$ (II-IV) or our theoretical value is included in the experimental error range (I). As an aside, we note that for the parameter sets I-III (without external force) also the simulated mean velocities coincide well with the experimental values with a maximum deviation of $10\%$ as shown in figure \ref{vmeanplot}. For illustrative purposes, we also plot $\qk$ plus $W$ for $\thetap=0.01$ which shows better agreement with the experimental data (but is not consistent with the range of $\thetap$ obtained in section \ref{model_param}).

Discrepancies between our theory and the experiment are visible in figures \ref{vmeanplot} and \ref{QWplot} where for parameter set V both the average velocity and the pseudo efficiency deviate significantly from the experimental values for $\fk=40\,\kt/d^2$ and $\thetap=0.1$. For $\dmu= 28.12 \kt$ corresponding to the data set V in figure \ref{QWplot}, the probe just reaches the potential minimum between consecutive jumps of the motor protein. Therefore on average at most $20\kt$ can be transferred to the spring leading to $\etaq\simeq0.7$, which is less than the experimental value. This discrepancy is most likely due to the fact that we have omitted substeps in our model. In the simulation, the average velocity then does not show saturation as one would expect it to result from the hydrolysis step \cite{yasu01} which should be experimentally observable at the higher concentrations used in \cite{toya10}. 

The confinement of $\thetap$ to the range $0.1\lesssim\thetap\lesssim 0.3$ implies on the one hand that the potential of mean force of the motor protein should be asymmetric and on the other hand that asymmetric potentials with a barrier state close to the initial state seem to enhance the ability of the motor protein to perform work on the spring, in accordance with \cite{schm08a}. If $\thetap$ was larger, $\etaq$ would decrease and the experimentally determined values of $\etaq$ would not be reached in the simulation. If $\thetap$ was smaller, $\etaq$ would approach the experimental values better, however, the distribution of the position of the probe just before a jump as shown in figure \ref{Hyjumps} would then no longer coincide with the experimentally observed distribution (see \cite{toya11a}). 

Information about the thermodynamic efficiency of the motor protein can be gained by applying an external force at the probe. In our simulation, the stall force is found to be $\fext=\dmu/d$, implying that the motor protein is able to convert the full $\dmu$ into extractable work without dissipation in accordance with the experiments performed in \cite{toya11}.

\section{Conclusion}

In summary, we have discussed a simple generic model which includes the elastic
linker between the probe particle and the molecular motor. Properties
of the motor become typically accessible only through the observation of the
motion of the probe. We have then focussed on discussing three types of
efficencies within this model using both simulations and a Gaussian
approximation to the stationary distribution for the distance between
motor and probe. The genuine thermodynamic efficiency is
non-zero only if an external force is applied to the probe. The Stokes
efficiency deviates from 1 due to the discrete nature of the motor
steps which become less relevant with increasing ATP concentration.
A pseudo efficiency measuring how much of the free energy of ATP hydrolysis
ends up in loading the elastic element can even become larger than 1
close to equilibrium and for a barrier state close to the initial state.

Applying this minimal model to recent experimental data for the $\feins$--ATPase
we find overall good agreement except for those parameters where especially the $\PI$ concentration is very small. In general,
one should consider ATP binding and ATP  hydrolysis as two separate
steps. Such a refinement as well as a splitting of the $120^{\circ}$ rotation
into two steps of $90^{\circ}$ and $30^{\circ}$ as experimentally observed using
much smaller probe particles does not pose new conceptual challenges
to the present framework and will be pursued elsewhere.

\appendix
\section{Equivalence of heat flow $\qdotk$ with the one inferred from the Harada--Sasa relation}

Experimentally, the heat flow caused by the probe has been inferred from measuring the autocorrelation function $\cv(\tau)=\mean{\dot{x}(t+\tau)\dot{x}(t)}-v^2$ and the linear response function 
\begin{equation}
\res(\tau)\equiv \frac{\delta \mean{\dot{x}(t+\tau)}}{\delta h(t)}\big{|}_{h=0}
\end{equation}
of the velocity of the probe to a small external perturbation $h(t)$ of the probe within the steady state \cite{toya10}. The heat flow is then given by an equality derived by Harada and Sasa \cite{hara05}
\begin{eqnarray}
 \qdoths&=\gamma v^2+\gamma\int_{-\infty}^\infty \frac{\mathrm{d}\omega}{2\pi} [\ftcv(\omega)-2\kt \,\mathrm{Re}\,(\ftres(\omega))] \nonumber \\
 &=\gamma v^2+\gamma[\cv(0)-2\kt \res(0)] \label{HS}\\
 &\equiv \dot{Q}_{\mathrm{S}}+ \dot{Q}_{\mathrm{V}},
\label{QSQV}
\end{eqnarray}
with $\ftcv$ and $\ftres$ being the Fourier transforms of $\cv$ and $\res$. Using a path weight approach described in \cite{seif09} applied to our system, the response function follows as
\begin{equation}
 \res(\tau)=\frac{1}{2\kt}\mean{\dot{x}(t+\tau)[\dot{x}(t)-(\fk\y(t)-\fext)/\gamma]}. 
\end{equation}
Inserting $\cv$ and this $\res$ into (\ref{HS}), one immediately finds
\begin{equation}
 \qdoths=\mean{\dot{x}(\fk y-\fext)}
\end{equation}
which is equal to $\qdotk$ in (\ref{heatflow}).

\section*{References}

\bibliographystyle{unsrt}
\bibliography{/home/public/papers-softbio/bibtex/refs}

\end{document}